\documentclass[aps,prd,twocolumn,superscriptaddress,showpacs,amsmath,amssymb]{revtex4}
\usepackage{graphicx,epsf}
\usepackage[]{latexsym}
\newcommand{\beq}{\begin{eqnarray}}
\newcommand{\eeq}{\end{eqnarray}}
\newcommand{\ud}{\mathrm{d}}

\newcommand{\lt}{\left(}
\newcommand{\rt}{\right)}
\newcommand{\lqu}{\left[}
\newcommand{\rqu}{\right]}
\newcommand{\vp}{\varphi}

\newcommand{\dota}{\dot{a}}
\newcommand{\ddota}{\ddot{a}}
\newcommand{\dotp}{\dot{\phi}}
\newcommand{\ddotp}{\ddot{\phi}}
\newcommand{\dotvp}{\dot{\varphi}}
\newcommand{\ddotvp}{\ddot{\varphi}}
\newcommand{\lag}{\mathcal{L}}
\newcommand{\vxp}{v - \xi \phi^2}
\newcommand{\vxpop}{v - \xi \hat{\phi}^2}
\newcommand{\wx}{W_\xi (\phi)}
\newcommand{\wxop}{\hat{W}_\xi}

\newcommand{\pop}{\hat{\phi}}
\newcommand{\vpop}{\hat{\varphi}}
\newcommand{\pa}{P_a}
\newcommand{\paop}{\hat{P}_a}
\newcommand{\pp}{P_\phi}
\newcommand{\ppop}{\hat{P}_\phi}

\newcommand{\pvpop}{\hat{P}_\varphi}
\newcommand{\ptau}{\partial_\tau}

\newcommand{\lp}{\ell_\text{P}}
\newcommand{\hub}{\mathcal{H}}

\newcommand{\hc}{\hat{H}_{0}}
\newcommand{\bi}{\hat{b}}
\newcommand{\bid}{\hat{b}^\dagger}

\newcommand{\es}{&\!\!=\!\!&}
\newcommand{\bra}[1]{\mbox{$\langle #1\! \mid$}}

\newcommand{\ket}[1]{\mbox{$\mid \!#1\rangle$}}

\newcommand{\pro}[2]{\mbox{$\langle #1 \mid #2\rangle$}}
\newcommand{\expec}[1]{\mbox{$\langle #1\rangle$}}

\begin{document}
\title{Inflation and the semiclassical dynamics of a conformal scalar field}
\author{G.~L.~Alberghi}
\email{alberghi@bo.infn.it}
\affiliation{Dipartimento di Fisica, Universit\`a di Bologna,
and I.N.F.N., Sezione di Bologna, via~Irnerio~46, 40126~Bologna, Italy}
\affiliation{Dipartimento di Astronomia, Universit\`a di Bologna,
via~Ranzani~1, 40127~Bologna, Italy.}
\author{C.~Appignani}
\email{appignani@bo.infn.it}
\affiliation{Dipartimento di Fisica, Universit\`a di Bologna,
and I.N.F.N., Sezione di Bologna, via~Irnerio~46, 40126~Bologna, Italy}
\author{R.~Casadio}
\email{casadio@bo.infn.it}
\affiliation{Dipartimento di Fisica, Universit\`a di Bologna,
and I.N.F.N., Sezione di Bologna, via~Irnerio~46, 40126~Bologna, Italy}
\author{F.~Sbis\`a}
\email{Fulvio.Sbisa@gmail.com}
\affiliation{Dipartimento di Fisica, Universit\`a di Bologna,
and I.N.F.N., Sezione di Bologna, via~Irnerio~46, 40126~Bologna, Italy}
\author{A.~Tronconi}
\email{tronconi@bo.infn.it}
\affiliation{Dipartimento di Fisica, Universit\`a di Bologna,
and I.N.F.N., Sezione di Bologna, via~Irnerio~46, 40126~Bologna, Italy}
\begin{abstract}
We derive the semiclassical evolution of massless conformally coupled scalar
matter in the de~Sitter space-time from the Born-Oppenheimer reduction of
the Wheeler-DeWitt equation.
We find a remarkable difference with respect to the minimally coupled case:
the effect of the quantum gravitational corrections do not depend on the momentum
of the scalar mode up to second order in the Planck length and, therefore, there are
no relevant corrections to the dispersion relation.
\end{abstract}
\maketitle
\section{Introduction}
Inflation~\cite{infla} has by now become the standard picture 
of the early Universe since it solves some of the problems of the standard
big-bang scenario and allows one to make testable predictions about the
spectrum of the cosmic microwave background radiation.
It also provides a window towards trans-Planckian physics~\cite{branden}
as it magnifies all quantum fluctuations and red-shifts originally
trans-Planckian frequencies down to the range of low energy
physics currently observed.
\par
In general, one expects that the standard results of quantum field theory
can be no more fully trusted on approaching the Planckian regime or, at least
that the dispersion relations of matter fields change for large wavenumber $k$.
Most attempts have tested the effects of dispersion relations $\omega=\omega(k)$
chosen {\em ad hoc\/}~\cite{branden}.
In Ref.~\cite{act_tp}, we have instead {\em derived\/} the dispersion relation
for a minimally coupled massless scalar field from a minisuperspace
action by lifting the principle of time-reparameterization
invariance to a quantum symmetry.
We then obtained an Hamiltonian constraint from which the
Born-Oppenheimer (BO) reduction~\cite{bv} allowed to properly
and unambiguously recover the semiclassical limit of quantum
field theory on a curved background starting from the
Wheeler-DeWitt (WDW) equation~\cite{dewitt}.
This procedure yields ``quantum fluctuation'' terms in the matter equation
whose effect on the power spectrum was derived for the simplest
model of de~Sitter inflation.
\par
In the present paper, we wish to extend the analysis to the case of a
generic coupling between the matter field and gravity and will also include
an inflaton field so as to consider a generic inflationary evolution, at least
for the more formal part.
We shall first show that it is possible to treat both the cosmological
scale factor $a$ and the inflaton semiclassically, and obtain the relevant
corrections for the matter field. 
We shall then specialise again to de~Sitter inflation, this time for
conformally coupled massless scalar matter and derive its effective
dynamics.
Quite remarkably, we shall see that no relevant corrections appear
at (relatively) large scale factor in the Schr\"odinger equation and the
dispersion relation remains unaffected, $\omega\simeq k/a$, to second
order in the Planck length.
Of course, it is well known that a conformally coupled scalar field can be
rescaled and, by making use of the conformal time, it decouples from
background gravity ({\em i.e.}, to zero order in the Planck length).
However, our approach allows us to show that, in a sense,
this decoupling extends to include quantum gravitational fluctuations
to second order in the Planck length.
The effect of such fluctuations is given in terms of complicated operators
acting on the scalar field state and the ``triviality'' of the final result is therefore
not at all obvious from the outset. 
\par
In the next Section, we shall briefly introduce the model and its classical
dynamics.
For the sake of generality, we shall treat the inflaton as an independent
degree of freedom minimally coupled to the cosmological scale factor
and also consider a generic coupling of the perturbation field to
gravity.  
The WDW equation in the BO decomposition will then
be analysed in Section~\ref{semicl} and its application to
the specific case of a conformally coupled perturbation in
de~Sitter inflation given in Section~\ref{dS}.
Finally, we shall comment on our findings in Section~\ref{concl}.
\par
We shall use units with $c=1$ and
$\kappa = 8\, \pi\, G_{\rm N}$.
\section{Classical model}
\label{classical}
Let us begin by briefly reviewing the classical dynamics for
the system in consideration.
More details can be found, for example, in Ref.~\cite{acg},
where however only one matter field was considered.
\par 
We start from the action
\beq
S
\es
\frac{1}{2} \int \ud^4 x\,\sqrt{-g} \left[
\left(\frac{1}{\kappa} - \xi\, \Phi^2 \right) R
-\left(\partial \Phi \right)^2 - \mu^2\,\Phi^2
\right.
\nonumber
\\
&&
\left.
\phantom{d x \sqrt{-g} \int \frac{1}{2\,\kappa}}\ \
- \left(\partial \Psi\right)^2 - m^2\, \Psi^2 - V(\Psi)
\right]
\ ,
\label{iniact}
\eeq
where $\Psi$ is the minimally coupled inflaton, $\Phi$
describes perturbations with a generic coupling $\xi$ to
gravity, $m$ and $\mu$ are the mass of
$\Psi$ and $\Phi$ respectively, $g$ is the determinant of
the metric and $V(\Psi)$ a potential for $\Psi$.
Given the FRW metric~\cite{mtw}
\beq
\ud s^2 = - N^2\,\ud t^2
+ a^2 \lt \frac{\ud r}{1- Kr^2}
+ r^2 \ud \Omega^2 \rt
\label{met}
\eeq
where $N = N(t)$ is the lapse function and $a=a(t)$ the
scale factor, the curvature scalar becomes
\beq
R = \frac{6}{N^2} \lt \frac{\ddot{a}}{a} - \frac{\dot{a}}{a}
\frac{\dot{N}}{N} + \frac{\dot{a}^2}{a^2} + N^2
\frac{K}{a^2} \rt
\label{curvsca}
\eeq
where $K=0,\pm 1$ is the spatial curvature and $\dot f=\partial_t f$.
\par
We are primarily interested in considering the field $\Phi$ as
a perturbation over the background driven by the inflaton $\Psi$.
Denoting with $\phi_k$ a given mode of $\Phi$ and with $\vp$
the homogenous mode of $\Psi$,
the above expression can therefore be written as the sum over
the actions for each mode of $\Phi$,
\begin{widetext}
\beq
S_{k}=
\frac{1}{2} \int N\,a^3\,\ud t
\left[
\frac{\dot{\phi_k}^2}{N^2} - \omega_k^2\, \phi_k^2
-\frac{6}{a^2} \lt v-\xi\,\phi_k^2 \rt
\lt \frac{\dot a^2}{N^2}-K\rt
+12\,\xi\,\frac{\phi_k\,\dot a\,\dot\phi_k}{a\,N^2}
+ \frac{\dot{\vp}^2}{N^2} - m^2 \vp^2 - V_\vp \right]
\ ,
\label{separact}
\eeq
\end{widetext}
in which we eliminated second derivatives of $a$ by integrating
by parts \cite{acg},
\beq
\omega^2_k = \frac{k^2}{a^2} + \mu^2
\ ,
\eeq
and $v= \mathcal{V}/ \kappa$ where $\mathcal{V}\, a^3$ is the
``volume of the universe''.
In the following, we shall just consider one mode at a time and
omit the corresponding index $k$ when this does
not cause confusion. 
\subsection{Lagrangean dynamics}
Varying the action and then setting $N=1$ (proper time
gauge $t=\tau$, so that $\dot f=\ptau f$) yields
\beq 
\frac{1}{a^3}\,\frac{\delta S}{\delta N}
\es
3  \lt v - \xi\, \phi^2\rt  \lt
\frac{\dot{a}^2}{a^2} + \frac{K}{a^2} \rt
-\frac{1}{2} \lt \dot{\phi}^2 + \omega^2\,\phi^2 \rt
\nonumber
\\
&&
- 6\, \xi\, \frac{\dot{a}}{a}\,\phi\,\dot{\phi}
-\frac{1}{2} \lt \dot{\vp}^2 + m^2\, \vp^2
+V_\vp \rt 
\nonumber
\\
\es
-\frac{H}{a^3}
=0
\label{hamcon}
\eeq
\beq
\frac{1}{a^2}\,\frac{\delta S}{\delta a}
\es
3\left\{ \lt v - \xi\, \phi^2\rt 
\lt 2\, \frac{\ddot{a}}{a} + \frac{\dota^2}{a^2} +
\frac{K}{a^2} \rt
- 4\, \xi\, \frac{\dota}{a}\, \phi\, \dotp
\right.
\nonumber
\\
&&
-2\, \xi
\lt \dotp^2 + \phi\, \ddotp \rt
+ \frac{1}{2} \lt \dotp^2 -\omega^2\, \phi^2 \rt
+ \frac{k^2\, \phi^2}{3\, a^2} 
\nonumber
\\
&&
\left.+
\frac{1}{2} \lt \dotvp^2 - m^2 \vp^2 - V_\vp \rt
\right\}
= 0
\eeq
\beq
\frac{1}{a^3}\,\frac{\delta S}{\delta \phi}
\es
-\lqu \ddotp
+ 3\,\frac{\dota}{a}\, \dotp + \omega^2\, \phi
+ 6\, \xi \lt 
\frac{\ddot{a}}{a} + \frac{\dota^2}{a^2}
+ \frac{K}{a^2} \rt
\phi \rqu
\nonumber
\\
\es
0
\eeq
\beq
\frac{1}{a^3}\,\frac{\delta S}{\delta \vp}
\es
- \lt \ddotvp + 3\,\frac{\dota}{a}\, \dotvp + m^2\, \vp
+ \frac{1}{2} V'_\vp \rt =
0
\ .
\eeq
It is also easy to verify that the Hamiltonian
constraint~(\ref{hamcon}) is conserved for any $\xi$,
\beq
\frac{\ud H}{\ud \tau} + \dota\, \frac{\delta S}{\delta a} + 
\dotp\, \frac{\delta S}{\delta \phi}
+ \dotvp\, \frac{\delta S}{\delta \vp} =0
\ .
\label{hamcons}
\eeq
\subsection{Hamiltonian dynamics}
The conjugate momenta are given by
\beq
\begin{array}{l}
P_N
=
\strut\displaystyle\frac{\partial \lag}{\partial \dot{N}} = 0,
\\
\\
P_a
=
\strut\displaystyle\frac{\partial \lag}{\partial \dota} = \frac{6\,a}{N}
\lqu \xi\, a\, \phi\, \dotp - \dota \lt v-\xi\, \phi^2 \rt\rqu
\\
\\
P_{\phi}
=
\strut\displaystyle\frac{\partial \lag}{\partial \dotp} = 
\frac{a^2}{N} \lt a\, \dotp + 6\, \xi\, \dota\, \phi \rt
\\
\\
P_{\vp}
=
\strut\displaystyle\frac{\partial \lag}{\partial \dotvp} = 
\frac{a^3\, \dotvp}{N}
\ ,
\end{array}
\eeq
and one can re-express the super-Hamiltonian as
\beq
H 
\es
- \frac{\lt a\, P_a - 6\,\xi\, \phi P_\phi \rt^2}{12\, a^3\, \wx} +
\frac{P_\phi^2}{2\, a^3} 
- 3\, K\, a \lt \vxp \rt
\nonumber
\\
&&
+ \frac{P_\vp^2}{2\, a^3}
+ \frac{a^3}{2}\lt \omega^2\, \phi^2 + m^2\, \vp^2 + V_\vp \rt
\ ,
\eeq
where we have again set $N=1$ and
\beq
\wx = v- \xi \lt 1- 6\, \xi\rt  \phi^2
\ .
\eeq
Of course, equivalent expressions of the super-Hamiltonian
can be implemented by applying canonical transformations.
For example, in order to quantise the system canonically,
it is more convenient to use $\bar{H}\equiv \wx\, H$
(see Ref.~\cite{acg} for the details), namely
\beq 
\bar{H}
\es
- \frac{\lt a\, P_a - 6\,\xi\, \phi\, P_\phi \rt^2}{12\, a^3}
\nonumber
\\
&&
+ \frac{\wx}{2\,a^3} \lqu
P_\phi^2 + P_\vp^2
- 6\, K\, a^4 \lt \vxp \rt
\right.
\nonumber
\\
&&
\left.
+a^6 \lt \omega^2\, \phi^2 + m^2\, \vp^2 + V_\vp \rt 
\rqu
\ .
\eeq
%
%
%
\section{Semiclassical model}
\label{semicl}
The quantization of the system is realized canonically, {\em i.e.}~by
replacing the classical variables and momenta with the respective
operators.
The Hamiltonian constraint $\bar{H} = 0$ then becomes the
WDW equation~\footnote{For a thorough discussion
of the minisuperspace,
the emergence of time and the WKB approximation, see
Refs. \cite{anderson}.}
\beq
\hat{H} \ket{\Psi} = 0
\ .
\eeq
A convenient way to treat this equation is to operate the
BO factorization~\cite{bv,acg,act_tp}
\beq
\Psi(a,\phi,\vp) = \psi(a)\,X(\phi, \vp;a)
\ .
\eeq
The operators $\hat{P_a}$, $\hat{P_\phi}$ and $\hat{P_\vp}$ are
defined as usual,
\beq
\hat{P_a}=-i\, \hbar\, \partial_a
\ ,
\quad \hat{P_\phi} = -i\, \hbar\, \partial_\phi
\ ,
\quad \hat{P_\vp} = -i\, \hbar\, \partial_\vp
\ ,
\eeq
and we also define the scalar product
\beq
\pro{X}{Y}=\int \ud \phi\, \ud \vp\,X^*(\phi,\vp;a)\,Y(\phi,\vp;a)
\ .
\eeq
\par
It is now convenient to factor out a geometrical phase and
redefine the wavefunctions as
\beq
\begin{array}{l}
\psi
\to
\tilde{\psi} =
e^{\strut\displaystyle{+\frac{i}{\hbar} \int \bra{X}\paop\ket{X}\, \ud a}} \psi
\\
\\
X
\to
\tilde{X} =
e^{\strut\displaystyle{- \frac{i}{\hbar} \int \bra{X}\paop\ket{X}\, \ud a}} X
\ ,
\end{array}
\eeq
from which it follows that  $\psi\, X= \tilde{\psi}\, \tilde{X}$ and
\beq
\bra{\tilde X}\paop\ket{\tilde X} = 0
\ .
\label{geophase}
\eeq
We shall omit tildes from now on.
\subsection{Gravitational equation}
We obtain the equation for the gravitational wavefunction
$\psi$ by contracting the WDW equation with $\bra{X}$,
and using Eq.~\eqref{geophase},
\beq
\bra{X} \hat{O}(\phi, \psi)\, \hat{P}_a \ket{X\, \psi} =
\lt \expec{\hat{O}\, \hat{\pa}}
+ \expec{ \hat{O} }\,
\hat{\pa} \rt \ket{ \psi }
\ ,
\eeq
where $\hat{O}$ is any operator and
$\expec{ \hat{O} } \equiv \bra{ X} \hat{O} \ket{ X }$.
In details,
\beq
&&
\frac{1}{2} \left\{-\frac{1}{6\, a^3} \lqu a^2\, \paop^2
- 12\, \xi\expec{ \pop\, \ppop}\, a\, \paop
+ 36\, \xi^2\, \expec{\pop^2\,\ppop^2} \rqu
\right.
\nonumber
\\
&&
+\frac{1}{a^3} \lqu \expec{ \wxop\, \ppop^2 }
+ \expec{ \wxop\,\pvpop^2} \rqu
-6\,K\,a\, \expec{ \wxop \lt\vxpop\rt}
\nonumber
\\
&&
\left.
\phantom{\frac{1}{2}}
+a^3 \lqu \omega^2 \expec{ \wxop\, \pop^2}
+ m^2\, \expec{\wxop\,\vpop^2}
+ \expec{ \wxop\, \hat V_\vp} \rqu
\right\} \ket{\psi}
\nonumber
\\
&&
=\frac{1}{2} \lt \frac{\expec{ \paop^2}}{6\,a}
- \frac{2\,\xi}{a^2} \expec{ \pop\, \ppop\, \paop} \rt
\ket{\psi}
\equiv
\hat{\Delta}^{(g)}\, \ket{\psi}. 
\eeq
\par
Assuming that the effect of $\hat\Delta^{(g)}$ is negligible and that
$\psi$ is peaked on a classical trajectory $a=a(\tau)$,
we can employ the WKB approximation, neglecting terms of order
$\hbar^2$ or higher, 
\beq
\psi \simeq \psi_{\textsc{wkb}} =  
e^{+\strut\displaystyle\frac{i}{\hbar} \lt S_0 + \hbar\, S_1 \rt }
\eeq
where $S_0$ and $S_1$ are implicitly defined by the
relations
\beq
\begin{array}{l}
\partial_a S_0
=
P_a
\\
\\
\partial_a S_1
=
f(a) \equiv \strut\displaystyle\frac{i}{2}\,\frac{\partial_a P_a}{P_a - z(a)}
\\
\\ 
z(a)
=
\strut\displaystyle\frac{6\, \xi}{a}\,\expec{ \pop\, \ppop}
\ ,
\end{array}
\eeq
whence 
\beq
\hat{\pa} \psi_{\textsc{wkb}} = 
\lqu P_a + \hbar\, f(a) \rqu \psi_{\textsc{wkb}}
\ .
\label{pawkb}
\eeq
This allows us to write the semiclassical Hamilton-Jacobi equation
for $a=a(\tau)$,
\beq
&&
3\,a
\lt \dota^2+ 
K\,\frac{ \expec{\wxop (\vxpop)}}{\expec{\wxop}^2}\rt
-\frac{\expec{\wxop\,\ppop^2}
+\expec{\wxop\,\pvpop^2}}{2\,a^3\,\expec{\wxop}^2} 
\nonumber
\\
&&
+ \frac{a^3}{2\,\expec{\wxop}^2} \lt \omega^2 \expec{\wxop\, \hat{\phi}^2}
+ m^2 \expec{ \wxop\,\hat{\vp}^2}
+ \expec{ \wxop\, \hat V_\vp} \rt
\nonumber
\\
&&
= - 3\,\xi^2\, \frac{\expec{ \hat{\phi}^2\, \ppop^2}
- \expec{ \hat{\phi}\, \hat{\pp}}^2}
{a^3 \expec{ \wxop}^2}
\equiv \Delta_\phi.
\eeq
in which we used the semiclassical expression for $\pa$,
\beq
\pa  = 6\, \frac{\xi}{a} \expec{ \hat{\phi}\, \ppop}
- 6\,a\, \dota \expec{ \wxop}
\ .
\label{semia}
\eeq
Neglecting $\Delta_\phi$, approximating the mean values of
products of operators with the products of their mean
values, replacing $\hat P_\phi$ and $\hat P_\vp$
with their classical expressions, and expanding $\wxop$,
one can finally recover the semiclassical gravitational
equation~\eqref{hamcon}.
\subsection{Matter equation}
The equation for the evolution of matter states is obtained
by computing
\beq
\hat{H} \ket{ X\, \psi}
- \ket{X} \bra{ X} \hat{H}\ket{X\, \psi}
=0
\ .
\label{matEq}
\eeq
Note that the procedure defined by this formula
implies that our matter equations will always contain
structures of the form $\hat O -\expec{\hat O}$ for
relevant operators $\hat O$, and this will turn out very
important, for example, for the operator ordering [see
the discussion after Eq.~\eqref{weyl}].
\par
We can introduce the (proper) time according to
\beq
\dota\, \paop = -i\, \hbar\, \dota\, \frac{\partial}{\partial a}
\equiv -i\, \hbar\, \frac{\partial}{\partial \tau}
\label{timedef}
\eeq
and, using \eqref{pawkb}, Eq. \eqref{matEq} yields
\beq
\lqu i\, \hbar \, \partial_\tau -
\lt \hat{H} - \expec{ \hat{H} }\rt \rqu \ket{ X } 
= \hat{\Delta}^{(m)} \ket{ X}
\ ,
\label{matt}
\eeq
where
\beq
\hat{H} = \hat{H}_\phi + \frac{\wxop}{\expec{ \wxop}}\,\hat{H}_\vp
\ ,
\eeq
and 
\beq
\hat{\Delta}^{(m)}
\es
-\frac{1}{12\,a\, \expec{\wxop}}\left\{
\paop^2 - \expec{ \paop^2}
\phantom{\frac{A}{B}}
\right.
\nonumber
\\
&&
-12\,\frac{\xi}{a}
\lqu \lt \pop\, \ppop - \expec{ \pop\, \ppop} \rt
\paop
-\expec{ \pop\, \ppop\, \paop} \rqu
\nonumber
\\
&&
+\frac{i\, \hbar}{a\, \dota}
\lt
\dota + \frac{a\, \ddota}{\dota}
+a\, \dota\, \frac{\partial_a\, \expec{ \wxop}}{\expec{ \wxop}}
\right.
\nonumber
\\
&&
\left.
\left.
+\xi\,\frac{\expec{ \pop\, \ppop}- a\,\partial_a \expec{ \pop\, \ppop}}{a^2\,\expec{ \wxop}}
\rt \paop
\right\}
\ .
\eeq
The two Hamiltonians are given by
\beq
\hat{H}_\phi
\es 
\frac{\wxop}{2\, \expec{ \wxop}} 
\lqu \frac{\ppop^2}{a^3} + \omega^2\, a^3\, \pop^2
- 6\,K\,a\lt \vxpop \rt
\rqu
\nonumber
\\
&&
- 6\,\xi\, \frac{\dota}{a}\, \pop\, \ppop
+ \frac{3\, \xi^2}{a^3\, \expec{ \wxop}} 
\lt  2\, \expec{ \pop\, \ppop}\, \pop\, \ppop
- \pop^2\, \ppop^2 \rt
\nonumber
\\
&&
+\frac{i\, \hbar \, \xi}{2\, a^3\, \dota\, \expec{ \wxop}^2}\,
\left[ \lt \dota + \frac{a\, \ddota}{\dota} \rt
\expec{ \wxop}
+a\, \dota\, \partial_a \expec{ \wxop }
\right.
\nonumber
\\
&&
\left.
+\frac{\xi}{a^2} \lt \expec{ \pop\, \ppop}
-a\,{\partial_a \expec{ \pop\, \ppop}} \rt
\right] \pop\, \ppop
\label{hchi}
\\
\nonumber
\\
\hat{H}_\vp
\es
\frac{1}{2} 
\lqu \frac{\pvpop^2}{a^3} + a^3 \lt m^2\,\vpop^2 + \hat V_\vp \rt
\rqu
\ .
\eeq
Upon neglecting $\hat{\Delta}^{(m)}$ and rescaling $\ket{ X}$
as
\beq
X \to \bar{X}
=
X\, e^{-\strut\displaystyle\frac{i}{\hbar}\, \int \expec{ \hat{H}}\,\ud \tau}
\ ,
\eeq
we recover the usual Schr\"odinger equation for the full matter wavefunction
\beq
i\, \hbar \, \ptau \ket{\bar X}
= \hat{H} \ket{ \bar{X}}
\ . 
\label{sch}
\eeq
\subsection{Separating the Schr\"odinger equation}
The scalar fields $\phi$ and $\vp$ still appear together
in Eq.~\eqref{sch}. 
However, if we write the matter wavefunction as the product
of a wavefunction for the inflaton and one for the
perturbation,
\beq
X(\phi,\vp;a) = \chi(\phi;a)\,\rho(\vp;a)
\ ,
\label{fact}
\eeq
we can derive two separate Schr\"odinger equations
for the wavefunctions $\chi$ and $\rho$
from Eq.~\eqref{matt}.
\par
To find an evolution equation for the inflaton,
we contract Eq.~\eqref{matt} with $\bra{ \chi}$ 
and obtain
\beq
\lqu i\, \hbar \, \ptau - 
\lt \hat{H}_\vp - \bra{\rho}\hat{H}_\vp\ket{\rho} \rt 
\rqu \ket{ \rho }
=\bra{ \chi } \hat{\Delta}^{(m)} \ket{ \chi\, \rho}
\ ,
\label{eqRho}
\eeq
where the right hand side is given by
\beq
\!\!\!\!\!\!
\bra{ \chi } \hat{\Delta}^{(m)}\ket{ \chi\, \rho}
\es 
-\frac{1}
{12\,a\,\expec{\wxop}} \lqu
\paop^2 - \bra{ \rho} \paop^2 \ket{ \rho}
\phantom{\frac{A}{B}}
\right.
\nonumber
\\
&&
+\frac{i\, \hbar}{a}
\lt 1 + \frac{a\,\ddota}{\dota^2}
+a\,\frac{\partial_a \expec{ \wxop}}{\expec{ \wxop}}
\right.
\nonumber
\\
&&
\left.
\left.
+\xi\, \frac{ \expec{ \pop\, \ppop}
- a\,\partial_a \expec{ \pop\, \ppop} }{\dot a\,a^2\,\expec{ \wxop}}
\rt
\paop
\rqu
\ket{\rho}
\ ,
\label{Drho}
\eeq
in which we used 
\beq
\lt \bra{ \chi } \pop\, \ppop\, \paop \ket{ \chi}
-\expec{ \pop\, \ppop\, \paop} \rt \ket{ \rho}
\es 0
\ .
\label{rhs2}
\eeq
We will assume that the effect of the quantum fluctuations~\eqref{Drho}
on the background can be neglected.
Hence, by rescaling $\rho$ as usual,
\beq
\rho\to\bar \rho=
\rho\,e^{-\strut\displaystyle\frac{i}{\hbar}\int \bra{\rho}\hat H_\vp\ket{\rho}\, \ud\tau}
\ ,
\eeq
Eq.~\eqref{eqRho} becomes a Schr\"odinger equation for the inflaton 
in which the field $\phi$ does not enter explicitly.
In the WKB approximation for $\rho$ (as well as $a$) one can therefore consider
solutions to the classical Einstein equations which are not affected by the
presence of the perturbation $\phi$
(the particular case of de~Sitter inflation will be analysed in the following
Section).
\par
We shall likewise obtain the evolution equation for the perturbing
field $\phi$.
Contracting equation \eqref{matt} with $\bra{\rho}$, we get the
following equation for  $\chi$,
\beq
&&
\!\!\!\!\!\!
\lqu i\, \hbar \, \ptau
- \hat{H}_\phi + \bra{\chi} \hat{H}_\phi\ket{\chi}
-\lt \frac{\wxop}{\expec{ \wxop}} -1 \rt
\bra{\rho} \hat{H}_\vp\ket{\rho} \rqu
\ket{\chi}
\nonumber
\\
&&
=
\bra{\rho } \hat{\Delta}^{(m)}\ket{\rho\, \chi}
\ ,
\label{chi}
\eeq
where
\beq
\bra{\rho } \hat{\Delta}^{(m)} \ket{\rho\,\chi}
\es
-\frac{1}{12\,a\,\expec{ \wxop }}
\left\{
\paop^2 - \bra{ \chi } \paop^2 \ket{ \chi}
\phantom{\frac{A}{B}}
\right.
\nonumber
\\
&&
-12\,\frac{\xi}{a}
\lqu \lt \pop\, \ppop - \expec{ \pop\, \ppop}\rt
\paop
-\expec{ \pop\, \ppop\, \paop } \rqu
\nonumber
\\
&&
+\frac{i\, \hbar}{a}
\lt
1 + \frac{a\, \ddota}{\dota^2}
+a\,\frac{\partial_a \expec{\wxop}}{\expec{\wxop}}
\right.
\nonumber
\\
&&
\left.
\left.
+
\xi\,\frac{ \expec{ \pop\, \ppop}
- a\,\partial_a \expec{ \pop\, \ppop}}{\dota\,a^2\,\expec{\wxop}}
\rt
\paop
\right\}
\ket{ \chi}
\ .
\label{nonappr}
\eeq
It is worth noting that the inflaton $\vp$ explicitly contributes
to the dynamics of $\chi$ whenever the operator $\wxop$
does not act trivially on $\chi$.
It is only in that case that one obtains  
\beq
\lt
i\, \hbar \, \ptau- \hat{H}_\phi\rt
\ket{\bar \chi}
=0
\ ,
\eeq
after neglecting $\bra{\rho } \hat{\Delta}^{(m)} \ket{\rho\,\chi}$
and rescaling
\beq
\chi\to\bar \chi=
\chi\,e^{-\strut\displaystyle\frac{i}{\hbar}\int \bra{\chi}\hat H_\phi\ket{\chi}\, \ud\tau}
\ .
\eeq
\section{De Sitter inflation}
\label{dS}
Since the minimally coupled case $\xi=0$ on the flat ($K=0$)
de~Sitter space-time has already been studied in details in
Ref.~\cite{act_tp}, we shall here apply the general formalism developed
so far to the conformal case $\xi = 1/6$ and $\mu=0$ with
\beq
a(\tau) = a_0\,e^{\hub\, \tau}
\ ,
\label{ads} 
\eeq
where $\hub$ is the Hubble constant and we set $a(0)=a_0$ for the arbitrary
value of the scale factor at proper time $\tau=0$~\footnote{This is not
necessarily the initial time of inflation.
In fact, in Appendix~\ref{sec:genxi}, we shall take initial conditions
at $\tau=\tau_0\to-\infty$.}.
We also require that the matter mode $k$ lies inside the de~Sitter
horizon, at least at the initial time $\tau=\tau_0$, so that
\beq
k>\hub\,a(\tau_0)
\ .
\label{dSh}
\eeq
Eq.~\eqref{nonappr} then simplifies considerably,
since $W_{1/6}=v$ and
\beq
\bra{\rho } \hat{\Delta}^{(m)} \ket{\rho\,\chi}
\es
- \frac{\lp^2}{12\,a\,\hbar}
\left\{
\paop^2 - \bra{ \chi } \paop^2 \ket{ \chi }
\phantom{\frac{A}{B}}
\right.
\nonumber
\\
&&
-\frac{2}{a}
\lqu \lt \pop\, \ppop - \bra{ \chi} \pop\, \ppop \ket{\chi} 
-i\,\hbar\rt
\paop
\right.
\nonumber
\\
&&
\left.
\left.
\phantom{\frac{A}{B}}\ \
- \bra{ \chi} \pop\, \ppop\, \paop \ket{\chi} \rqu
\right\}
\ket{ \chi }
\ ,
\label{appr}
\eeq
where $\lp=v^{-1/2}$ is the Planck length.
All these contributions, once evaluated on an invariant
eigenstate (see next Subsection), will turn out to be of order
$\hbar\,\lp^2$.
\par
The Hamiltonian is now given by
\beq
\hat{H}_\phi
\es
\frac{1}{2} \lt \frac{\ppop^2}{a^3} 
+ \omega^2\, a^3\, \pop^2 \rt - \hub\, \pop\, \ppop 
+ \lp^2\, \hat{H}_{p}
\nonumber
\\
\es
\hat{H}_{0}+ \lp^2\, \hat{H}_{p}
\ ,
\label{hchiappr}
\eeq
where we have explicitly separated the perturbing
Hamiltonian 
\beq
\hat{H}_{p}
=
- \frac{\pop^2\, \ppop^2}{12\, a^3\, \hbar}
+ \frac{\expec{ \pop\, \ppop}\, \pop\, \ppop}{6\, a^3\, \hbar} 
+\frac{i\,\pop\, \ppop}{6\, a^3}
\ .
\eeq
A final simplification can be obtained by re-phasing $\chi$
according to
\beq
\chi \to \chi_s=
\chi\, \ e^{-\strut\displaystyle\frac{i}{\hbar}\, \int
\bra{\chi} \hat{H}_{0}\ket{\chi}\,\ud\tau}
\ ,
\eeq
which yields
\beq
\lt i\, \hbar \, \ptau -\hat{H}_{0} \rt \ket{\chi_s}
=
\hat{\Delta}_s \ket{\chi_s}
\ ,
\label{hchibar}
\eeq
where
\beq
\hat{\Delta}_s
\es
\frac{\lp^2}{6\,\hbar\,a^3}
\left\{
\frac{1}{2} \lt 
\expec{ \pop^2\, \ppop^2}_s 
-\pop^2\, \ppop^2\rt
\right.
\nonumber
\\
&&
+ \lt \frac{\expec{ \hc }_s}{\hub} + \expec{ \pop\, \ppop}_s
+i\, \hbar \rt \lt \pop\, \ppop - \expec{\pop\, \ppop}_s \rt
\nonumber
\\
&&
-\frac{i\, \hbar}{\hub} \lt \pop\, \ppop\, \ptau
- \expec{ \pop\,\ppop\, \ptau}_s \rt
+ \frac{\hbar^2}{2\, \hub^2} \lt
\ptau^2 - \expec{ \ptau^2}_s \rt
\nonumber
\\
&&
+ 
\frac{i\,\hbar}{\hub} \lqu \frac{\expec{ \hc }_s}{\hub}
+ \frac{i}{2}\,\hbar\rqu
\lt \ptau - \expec{ \ptau}_s \rt
\nonumber
\\
&&
\left.
+ \frac{1}{\hub} \lqu
\expec{ \pop\, \ppop}_s + i\, \hbar \rqu
\lt i\, \hbar\, \ptau -\expec{\hc}_s \rt
\right\}
\ ,
\label{deltachi}
\eeq
with $\expec{ \hat{O}}_s =\bra{\chi_s} \hat{O} \ket{\chi_s}$
for any operator $\hat{O}$.
\par
So far no particular operator ordering was chosen for
$\pop$ and $\ppop$.
We shall now choose the Weyl ordering by symmetrising
$\hat{\Delta}_s$ in $\pop$ and $\ppop$~\cite{weyl} according to
\beq
\begin{array}{l}
2\,\pop\,\ppop
\to
\pop\, \ppop + \ppop\,\pop=
2\,\pop\, \ppop - {i\, \hbar}
\\
\\
\pop^2\, \ppop^2 \to 
\pop^2\, \ppop^2 - 2\,i\,\hbar\,\pop\,\ppop
-{\hbar^2}/{2}
\ .
\end{array}
\label{weyl}
\eeq
Any other ordering would eventually result in different c-number
terms (proportional to $\hbar^2$) which do not appear in $\hat\Delta_s$
due to the general form $\hat \Delta_s\sim \hat O -\expec{\hat O}_s$
[as we noted after Eq.~\eqref{matEq}].
Substituting in Eq.~\eqref{hchibar}, after some algebra, one obtains
\beq
\lt 1
-\frac{\lp^2\,\Sigma}{6\,\hub\,a^3}
\rt
\lt i\, \hbar \, \ptau -\hc \rt \ket{\chi_s}
=
\frac{\lp^2 \, \hat{\Delta}}{6\,\hub\,a^3} \ket{\chi_s}
\ ,
\label{schfinal}
\eeq
where
\beq
\Sigma=\frac{i}{2} + \frac{\expec{ \pop\, \ppop}_s}{\hbar}
\ ,
\eeq
and
\beq
\hat{\Delta}
\es
\frac{\hub}{2\,\hbar} \lt 
\expec{ \pop^2\, \ppop^2}_s
-\pop^2\, \ppop^2\rt
\nonumber
\\
&&
+\frac{\hub}{\hbar} \lt \frac{\expec{ \hc}_s}{\hub} + \expec{ \pop\, \ppop}_s
+\frac{3}{2}\,i\, \hbar \rt
\lt \pop\, \ppop - \expec{ \pop\, \ppop}_s\rt
\nonumber
\\
&&
-i \lt \pop\, \ppop\, \ptau
-\expec{\pop\,\ppop\, \ptau}_s \rt
+ \frac{\hbar}{2\,\hub} \lt
\ptau^2 - \expec{\ptau^2}_s \rt
\nonumber
\\
&&
+i \lt  \frac{\expec{ \hc}_s}{\hub} + i\,\hbar\rt
\lt \ptau - \expec{ \ptau}_s\rt
\nonumber
\\
&&
+\lt \frac{\expec{ \pop\, \ppop}_s}{\hbar} + \frac{i}{2} \rt
\lt \hc - \expec{ \hc }_s\rt
\ .
\label{perturb}
\eeq
These are the expressions which we shall estimate
in the following.
\subsection{Perturbative analysis}
We now proceed to solve the matter equation~\eqref{schfinal}
perturbatively (in $\lp$).
As can be seen by looking at Eq.~\eqref{schfinal} and \eqref{perturb}
[see also the discussion following
Eq.~\eqref{hambbb}], the terms denoted by $\Sigma$ and
$\Delta$ represent a perturbation to the usual Schr\"odinger
equation only when
\beq
\epsilon= \frac{\lp^2}{6\,\hub\, a^3} \ll 1
\ ,
\eeq
so, strictly speaking, it is only for (relatively) late times such
that~\footnote{We assume that during inflation
$\hub\sim 10^{-3}\,\lp^{-1}$~\cite{infla}.}
\beq
a(\tau)\gg
\lt \frac{\lp^2}{6\,\hub}\rt^{1/3}
\sim
\lp
\ ,
\eeq
that the perturbative treatment makes sense.
In what follows, we will then assume that this relation holds
and neglect terms proportional to $\epsilon^2$, so that
\beq
\frac{\epsilon\,\hat{\Delta}}{1-\epsilon\,\Sigma}\simeq
\epsilon\, \hat{\Delta}
\ .
\eeq
However, we shall also discuss what might happen outside of this 
regime.
\par
To zero order in $\epsilon$, the matter equation is just~\footnote{This $O(\lp^0)$
equation reproduces the well-known result that conformally coupled scalar fields
evolve freely on a cosmological background.}
\beq
\lt i\, \hbar\, \ptau - \hc \rt \ket{\chi_s} = 0
\ ,
\label{mattzero}
\eeq
with an Hamiltonian of the form
\beq
\hc
\es
 \frac{1}{2} \lt \frac{\ppop^2}{a^3} 
+ \omega^2\, a^3\, \pop^2 \rt
- \frac{\hub}{2} \lt \pop\, \ppop + \ppop\, \pop \rt
\nonumber
\\
\es
\frac{\hbar\,k}{a} 	
\lt \bid\, \bi +\frac{1}{2} \rt
-\frac{i}{2}\,\hbar\,\hub \lqu (\bid)^2-\bi^2  \rqu
\ ,
\label{hambbb}
\eeq
in which the invariant operators $\bi$ and $\bid$ are defined
in Appendix~\ref{sec:genxi}.
Note that  the squeezing term containing $(\bid)^2$ and $\bi^2$
vanishes for vanishing $\hub$, although this limit is critical for
our treatment in that the definition of time~\eqref{timedef}
loses its meaning for $\dot a\to 0$~\footnote{More appropriately,
one should then use a different degree of freedom to introduce
the time.}.
In particular, on assuming that matter modes are generated in a
Bunch-Davies vacuum [tantamount to our condition~\eqref{B-D}]
and using Eqs.~\eqref{refstart}--\eqref{refend},
the effect of the perturbation $\hat{\Delta}$
given in Eq.~\eqref{perturb} on invariant
eigenstates $\ket{\chi_s}=\ket{n}$ can be easily estimated as
\beq
\epsilon\, \hat{\Delta}
\simeq
\epsilon\,\frac{\hbar\,\hub}{4}
\lqu
\lt  \bid\rt^2
-\bi^2
\rqu
=
\frac{\lp^2\, \hbar}{24\,a^3}
\lqu
\lt  \bid\rt^2
-\bi^2
\rqu
\ .
\label{epsll1}
\eeq
\par
Even if the regime $\epsilon\gtrsim 1$ is clearly
non-perturbative, one could try to examine in a qualitative
way what happens assuming that Eqs.~\eqref{refstart}--\eqref{refend}
still hold approximately.
One then finds
\beq
\frac{\epsilon\, \hat{\Delta}}{1-\epsilon\,\Sigma}
\simeq
\frac{1+i\,\epsilon}{1+\epsilon^2}\,
\epsilon\,
\frac{\hbar\,\hub}{4}
\lqu
\lt  \bid\rt^2
-\bi^2
\rqu
\ .
\eeq
For $\epsilon\sim 1$, this expression then yields
\beq
\frac{\epsilon\, \hat{\Delta}}{1-\epsilon\,\Sigma}
\simeq
\frac{1+i}{8}\,
\hbar\,\hub
\lqu
\lt  \bid\rt^2
-\bi^2
\rqu
\ ,
\label{eps1}
\eeq
and, for $\epsilon\gg 1$,
\beq
\frac{\epsilon\, \hat{\Delta}}{1-\epsilon\,\Sigma}
\simeq
\frac{i}{4}\,
\hbar\,\hub
\lqu
\lt  \bid\rt^2
-\bi^2
\rqu
\ .
\label{epsgg1}
\eeq
It is remarkable that, in  all regimes, the correction amounts to a
mere renormalization of the coefficient of the squeezing term independent
of $k$, so that the effective Schr\"odinger equation can be written as
\beq
\lt i\,\hbar\, \ptau  -
\hat H_{\rm{new}}\rt
\ket{\chi_s} =0
\ ,
\label{schnew}
\eeq
where
\beq
\hat H_{\rm{new}}
=
\frac{\hbar\,k}{a} 	
\lt \bid\, \bi +\frac{1}{2} \rt
-\frac{i}{2}\,\beta\,\hbar\,\hub \lqu (\bid)^2-\bi^2  \rqu
\ ,
\label{hnew}
\eeq
and
\beq
\beta=
\left\{
\begin{array}{ll}
1+\strut\displaystyle\frac{i\,\lp^2}{12\,\hub\,a^3}
&
\quad
{\rm for}
\quad
\epsilon\ll 1
\\
\\
\strut\displaystyle\frac{3+i}{4}
&
\quad
{\rm for}
\quad
\epsilon\sim 1
\\
\\
\strut\displaystyle\frac{1}{2}
&
\quad
{\rm for}
\quad
\epsilon\gg 1
\end{array}
\right.
\eeq
Except for the case $\epsilon\gg 1$, the operator $\hat H_{\rm new}$
is however not Hermitian, despite the fact that the approach followed ensures
that the evolution of the system remains unitary~\cite{bertoni}.
Note also that the squeezing term remains negligible for
large momenta, that is $a\,\hub/k\ll 1$.
\subsection{Late time dispersion relation}
The case $\epsilon\ll 1$ is similar to $\lp^2\,\hub^2=\delta^2\ll 1$ in
Ref.~\cite{act_tp}.
One can therefore employ the same kind of perturbative expansion,
\beq
\ket{\chi_s}=\ket{n_s}\simeq
\left(\hat 1+\delta^2\,\hat R_n\right) \ket{n}
\ ,
\label{chis1}
\eeq
where $\hat R_n$ must then satisfy
\beq
\left(i\,\hbar\,\ptau\hat R_n-\left[\hat H_0,\hat R_n\right]\right)
\ket{n}
=\frac{\epsilon}{\delta^2}\,\hat\Delta\ket{n}
\ .
\eeq
This implies that
\beq
\hat R_n=r\,\bi^2+r^*\,(\bid)^2
\ ,
\eeq
where
\beq
i\,a\,\dot r+2\,k\,r=-\frac{1}{24\,\hub^2\,a^2}
\ ,
\label{req}
\eeq
whose solution [assuming $r(\tau\to\infty)=0$] is given by
\beq
r(\tau)
=
\frac{1}{96\,k^3}\left[
1+\frac{2\,i\,k}{\hub\,a}
\left(1
-\frac{i\,k}{\hub\,a}
\right)
-e^{-\strut\displaystyle\frac{2\,i\,k}{\hub\,a}}
\right]
\ ,
\eeq
from which it is again clear that $\hat R_n$ is not anti-Hermitian
[$a=a(\tau)$ is given in Eq.~\eqref{ads}].
\par
In order to determine a (modified) dispersion relation for the
mode $k$, we first need to determine an effective Hermitian
Hamiltonian $\hat{H}_{\rm{eff}}$ so that the states $\ket{\bar\chi_s}$
evolved by it produce the same expectation values for any observables
$\hat X$ as those given by $\ket{\chi_s}$ in Eq.~\eqref{chis1},
\beq
\bra{\bar\chi_s} \hat X \ket{\bar\chi_s}=
\bra{\chi_s} \hat X \ket{\chi_s}
\ .
\label{cher}
\eeq
To first order in $\epsilon$, we can write
\beq
\hat{H}_{\rm{eff}}=i\,\hbar\,\lt\ptau \hat U\rt \hat U^{-1}
\ ,
\eeq
where the complete propagator $\hat U$ is given by
\beq
\hat U\simeq \lt \hat 1+i\,\delta^2\,\hat H_n\rt \hat U_0
\ ,
\eeq
with $\hat U_0$ the propagator
for the Schr\"odinger equation~\eqref{mattzero}, so that
\beq
\ket{\bar \chi_s}=\ket{\bar n_s}=\hat U\,\ket{n}
\simeq \lt \hat 1+i\,\delta^2\,\hat H_n\rt \ket{n}
\ .
\eeq
After some algebra, one finds
\beq
\hat{H}_{\rm{eff}}\simeq
\hat H_0
-\delta^2\lt \hbar\, \ptau \hat H_n
+i\lqu \hat H_0,\hat H_n\rqu\rt
\ .
\eeq
The Hermitian operator $\hat H_n$ can be finally determined
by imposing the condition~\eqref{cher} for $\hat X=\pop^2$,
$\ppop^2$ and $\{\pop,\ppop\}$, which yields
\beq
\hat H_n=i\,\frac{n^2+n+1}{2\,n+1}\,
\lqu r\,\bi^2-r^*\,(\bid)^2\rqu
\ ,
\eeq
where $r=r(\tau)$ is the same solution to Eq.~\eqref{req}.
Putting all the pieces together, the effective Hamiltonian
is then given by
\beq
\hat{H}_{\rm{eff}}
&\!\!\simeq\!\!&
\hat H_0
+\frac{\lp^2\, \hbar}{24\,a^3} \lqu \lt\bid\rt^2+\bi^2\rqu
\nonumber
\\
\es
\frac{1}{2}
\lt\frac{\ppop^2}{\mu}+\mu\,\omega_{\rm{eff}}^2\,\pop^2\rt
+\gamma\lt\pop\,\ppop+\ppop\,\pop\rt
\ ,
\label{Heff}
\eeq
with effective mass and frequency
\beq
\mu(\xi=1/6)
&\!\!\simeq\!\!&
a^3\lt 1+\frac{\lp^2}{12\,k\,a^2}\rt
\\
\omega_{\rm eff}(\xi=1/6)
&\!\!\simeq\!\!&
\omega
\ ,
\eeq
and
\beq
\gamma(\xi=1/6)
=-\frac{\hub}{2}
\ .
\eeq
To make the comparison easier, we report here the results
for the minimally coupled case~\footnote{Note that there are typos in
the numerical coefficients displayed in Ref.~\cite{act_tp}.},
\beq
\mu(\xi=0)
&\!\!\simeq\!\!&
a^3
\lt1+ \frac{3\,\lp^2}{k\,a^2} \rt
\\
\omega_{\rm{eff}}(\xi=0)
&\!\!\simeq\!\!&
\omega 
\lt 1+ \frac{3\,\lp^2\,\hub^2}{4\,k^3} \rt
\\
\gamma(\xi=0)
&\!\!\simeq\!\!&
\frac{3\,\lp^2\,\hub}{4\,k\,a^2}
\ .
\eeq
The most evident difference is therefore that the effective
frequency $\omega_{\rm eff}(\xi=1/6)\simeq\omega$ to
first order in $\lp^2$.
The effective mass is changed in both cases by terms proportional
to $1/k$, which can be interpreted as a different ``effective'' cosmological
scale factor $a_{\rm eff}=\mu^{1/3}(k)$ for each matter mode.
We remark that matter modes are generated inside the Hubble horizon,
so that no infrared divergence occurs in the above expressions
by virtue of Eq.~\eqref{dSh}.
\section{Conclusions}
\label{concl}
We have analysed the dynamics of a scalar field generically coupled to
gravity driven by an inflaton following the approach of Refs.~\cite{bv,acg}.
After showing under which conditions one can treat the cosmological
scale factor and the inflaton as (coupled) classical degrees of freedom,
we specialized to massless conformally coupled matter in de~Sitter inflation.
\par 
The main conclusion is that, at least in the perturbative regime $\epsilon\ll 1$
corresponding to a relatively large cosmological scale factor (compared to the
Planck length), quantum gravitational corrections
in the Schr\"odinger equation are negligible, since one can introduce an
effective Hamiltonian~\eqref{Heff} with the same effective
frequency and squeezing term (to order $\lp^2$) that appear in the usual
equation~\eqref{mattzero} obtained from quantum field theory
on that classical background.
This result can be viewed as an extension to order $\lp^2$
of the well-known fact that conformally
coupled scalar fields evolve freely on a cosmological background
and is strikingly different from what we obtained for a
minimally coupled scalar field in Ref.~\cite{act_tp}, for which
both the effective frequency and squeezing term instead contain
corrections starting at order $\lp^2$.
\appendix
\section{Matter states for general coupling}
\label{sec:genxi}
The unperturbed Eq.~\eqref{mattzero}
can be solved for a general value of $\xi$ provided
\beq
\frac{\wxop}{\expec{\wxop}} \simeq 1
\ ,
\label{W/W}
\eeq
so that
the Hamiltonian $\hc$ can be expressed as
\beq
\hc=
\hbar\, \omega_\xi \lt\hat{a}^\dagger_\xi\,\hat{a}_\xi +
\frac{1}{2} \rt
\ ,
\label{hamwxi}
\eeq
where
\beq
\omega_\xi^2
= \omega^{2} + 6\, \xi \lt
\frac{K}{a^2} - 6\, \xi\, \frac{\dota^{\,2}}{a^2} \rt 
\label{ome1}
\eeq
and $[\hat{a}_\xi,\hat{a}_\xi^\dagger]=1$ with
\beq
\hat{a}_\xi
\es
\sqrt{\frac{a^3\, \omega_\xi}{2\, \hbar}}
\lqu \hat{\phi} + \frac{i}{\omega_\xi} \lt
\frac{\hat{P}_{\phi}}{a^3} - 6\, \xi\, \frac{\dota}{a}\,
\hat{\phi} \rt \rqu
\ .
\label{axi}
\eeq
\par
For this Hamiltonian, the relevant invariant operators~\cite{lewis,acg,act_tp}
satisfying $[ \hat{b}_\xi, \hat{b}_\xi^\dagger]=1$ are given by
\beq
\hat{b}_\xi
\es
i \sqrt{\frac{a^3 \sigma^2}{2\, \hbar}}
\left\{
\frac{\hat{P}_{\phi}}{a^3}
-\lqu
\frac{i}{\sigma^2}+
\frac{\dot{\sigma}}{\sigma}
+6\,\frac{\dota}{a}\lt\xi-\frac{1}{4}\rt\rqu
\hat{\phi}\right\} 
\ ,
\label{bgen}
\eeq
and its Hermitian conjugate.
The function $\sigma=\sigma(\tau)$ must be a solution to the 
Pinney equation 
\beq
\ddot{\sigma} + \Omega^2\, \sigma = \frac{1}{\sigma^3}
\ ,
\label{Pinney}
\eeq
with
\beq
\Omega^2 = \omega_\xi^2
+ \lt 12\, \xi -\frac{3}{4} \rt
\frac{\dota^2}{a^2}
+ \lt 6\,\xi - \frac{3}{2} \rt
\frac{\ddot{a}}{a}
\ .
\eeq
The eigenstates of the invariant number operator, 
\beq
\hat{n}_b\ket{n}=\hat{b}_\xi^\dagger\, \hat{b}_\xi \ket{n}=
n\,\ket{ n}
\ ,
\eeq
then provide us with the exact solutions we are looking for,
namely 
\beq
\ket{n} =
\frac{e^{-i\,n\,\Theta(\tau)}}{\sqrt{n!}}\lt \hat{b}_\xi^\dagger\rt^n
\ket{0}
\ ,
\eeq
where $\hat{b}_\xi\ket{0}=0$ and
\beq
\Theta = \int^\tau\frac{\ud t}{\sigma^2(t)}
\ .
\eeq
\par
For a de~Sitter space-time, with 
$\dota^2/a^2 = \ddot{a}/a = \hub^2$,
we then have
\beq
\Omega^2 = A \, e^{-2\,\hub\, \tau} - B
\ ,
\eeq
where
\beq
A=k^2+6\,\xi\, K
\ ,\quad
B=36\lt\xi- 1/4 \rt^2 \hub^2-\mu^2
\ .
\eeq
\par
The general solution to Eq.~\eqref{Pinney} can be written as
\beq
\sigma = \frac{1}{w} 
\sqrt{c_1\, h_1^2 + c_2\, h_2^2
+ 2\,\sqrt{c_1\, c_2 -w^2} \, h_1\, h_2}
\ ,
\label{sigma}
\eeq
where $h_1$ and $h_2$ must be any two linear independent
solutions to the  auxiliary homogeneous equation
\beq
\ddot{\sigma} + \Omega^2\, \sigma = 0
\ ,
\label{aux}
\eeq
$w=\dot{h}_1\, h_2 - h_1\, \dot{h}_2$ is their Wronskian
and $c_1$, $c_2$ are coefficients to be fixed
by appropriate initial conditions at $\tau=\tau_0$.
For example, a common requirement is that
the invariant number operator and the Hamiltonian
are proportional at $\tau=\tau_0=-\infty$~\cite{acg,act_tp}
\beq
\hat{n}_b(\tau=-\infty) \propto \hc(\tau=-\infty)
\ .
\label{B-D}
\eeq
This amounts to requiring that 
\beq
\hat{a}^\dagger_\xi\,\hat{a}_\xi
\propto
\hat{b}^\dagger_\xi\, \hat{b}_\xi
\ ,
\quad
{\rm at}\  \tau=-\infty
\ .
\eeq
It can be shown that this uniquely implies that 
\beq
\hat{a}_\xi \propto \hat{b}_\xi
\  ,
\quad
\hat{a}^\dagger_\xi \propto \hat{b}^\dagger_\xi
\ ,
\quad
{\rm at}\  \tau=-\infty
\ ,
\eeq
and from these relations one finds the appropriate initial
conditions for $\sigma$ as
\beq
\begin{array}{l}
\sigma (\tau_0=-\infty)
=
\strut\displaystyle\frac{1}{\sqrt{\omega_\xi(-\infty)}}
\\
\\
\dot{\sigma} (\tau_0=-\infty) 
=
\strut\displaystyle\frac{3}{2}\, \hub\, \sigma (-\infty) =
\frac{3\,\hub}{2\,\sqrt{\omega_\xi(-\infty)}}
\ .
\end{array}
\label{incond}
\eeq
It also follows that $\ket{0}$ is then the usual Bunch-Davies
vacuum of minimum energy at $\tau=\tau_0$.
\par
A convenient pair of solutions $h_1$ and $h_2$
is given in terms of the Hankel functions
\beq
\begin{array}{l}
h_1(\tau; \nu)
=
i\, H^{(1)}_\nu(z)
=
i\,J_\nu(z) - N_\nu(z)
\\
\\
h_2(\tau; \nu)
=
i\,H^{(2)}_\nu(z) 
=
i\,J_\nu(z) + N_\nu(z)
\ ,
\end{array}
\eeq
where $H$, $J$ and $N$ are the Hankel, Bessel and Neumann
functions respectively,
\beq
\nu = \frac{\sqrt{B}}{\hub}
\ , 
\quad
z=\frac{\sqrt{A}}{\hub} \, e^{-\hub t}
\ ,
\eeq
and $w = i\, 4\, \hub / \pi$~\footnote{We remark that using
the Bessel functions as solutions $h_1$ and $h_2$ some cases,
such as $\xi=1/4$ and $\mu=0$ cannot be analysed because the
Wronskian $W$
becomes singular and the solution for $\sigma$ does not
exist or is not unique.}.
\par
We can now use the large $z$ expressions~\footnote{More precisely,
these expressions hold for
$z \gg |\nu^2 -1/4| = \left| \pm \lqu(6\,\xi -3/2)^2 -
\frac{\mu^2}{\hub^2} \rqu - \frac{1}{4} \right|$
and therefore, essentially, when $z \gg \frac{\mu^2}{\hub^2}$
and $z \gg 1$.} 
of the  functions $h_1$ and $h_2$,
\beq
h_1
&\!\!\simeq\!\! &
i\,e^{\frac{\hub\, \tau}{2}} 
\sqrt{\frac{\hub}{\sqrt{A}}} 
\sqrt{\frac{2}{\pi}}
\lqu
 \cos \lt\frac{\hub}{\sqrt{A}} e^{-\hub\,\tau}-
\frac{\pi}{4}-\frac{\sqrt{B}\,\pi}{2\,\hub}
\rt
\right.
\nonumber
\\
&&
\phantom{i\,e^{\frac{\hub\, \tau}{2}} 
\sqrt{\frac{\hub}{\sqrt{A}}} 
\sqrt{\frac{2}{\pi}}}
\left.
+i\,\sin \lt\frac{\hub}{\sqrt{A}} e^{-\hub\, \tau}-
\frac{\pi}{4}-\frac{\sqrt{B}\, \pi }{2\, \hub}\rt
\rqu
\nonumber
\\
\es
\sqrt{\frac{\hub}{\sqrt{\pi A}}}\, (i-1) \, 
e^{-i \lt \frac{\hub}{\sqrt{A}} e^{-\hub\,\tau} - 
\frac{\sqrt{B}\, \pi }{2\, \hub}\rt} \,e^\frac{\hub\, \tau}{2} 
\\
h_2
&\!\!\simeq\!\! &
i\,e^{\frac{\hub\, \tau}{2}} 
\sqrt{\frac{\hub}{\sqrt{A}}} 
\sqrt{\frac{2}{\pi }} \lqu  \cos 
\lt\frac{\hub}{\sqrt{A}} e^{-\hub\, \tau}-
\frac{\pi}{4}-\frac{\sqrt{B}\, \pi }{2\, \hub}\rt
\right.
\nonumber
\\
&&
\left.
\phantom{i\,e^{\frac{\hub\, \tau}{2}} 
\sqrt{\frac{\hub}{\sqrt{A}}} 
\sqrt{\frac{2}{\pi}}}
-i\, \sin \lt\frac{\hub}{\sqrt{A}} e^{-\hub\, \tau}-
\frac{\pi}{4}-\frac{\sqrt{B}\, \pi }{2\, \hub}\rt
\rqu
\nonumber
\\
\es
\sqrt{\frac{\hub}{\sqrt{\pi A}}}\; (1+i) \; 
e^{i \lt \frac{\hub}{\sqrt{A}} e^{-\hub t}- 
\frac{\sqrt{B} \pi }{2 \hub}\rt} \,e^\frac{\hub t}{2}
\ . 
\eeq
Then, on considering that 
\beq
\omega_\xi \simeq \sqrt{A} \, e^{-\hub\, \tau}
\ ,
\quad 
{\rm for}\ \tau \to -\infty
\ ,
\eeq
and imposing the initial conditions~\eqref{incond}
on Eq.~\eqref{Pinney}, one finds that 
$c_1$ and $c_2$ as function of the initial time $\tau_0$
must decay exponentially for  $\tau_0 \to -\infty$,
\beq
c_1
\es
\omega_\xi  
\,h_2^2
+\frac{\lt2\, \dot{h}_2 - 3\, \hub\, h_2 \rt^2}{4\, \omega_\xi}
\\
&\!\!\simeq\!\!&
- \frac{6\, \hub^2}{A\,\pi}\, 
e^{\hub\, \tau_0+i \lt \frac{2\,\hub}{\sqrt{A}}\,e^{-\hub\, \tau_0}
- \frac{\sqrt{B}}{\hub}\, \pi \rt}
\lt \sqrt{A} - \frac{3}{4}\, i\, \hub\, e^{\hub\, \tau_0}
\rt
\nonumber
\\
c_2 
\es
\omega_\xi\,h_1^2
+\frac{\lt 2\,\dot{h}_1 - 3 \hub h_1 \rt^2}{4 \omega_\xi}
\\
&\!\!\simeq\!\!&
- \frac{6\, \hub^2}{A\,\pi}\,
e^{\hub\, \tau_0+i\lt \frac{\sqrt{B}}{\hub}\,\pi
- \frac{2\,\hub}{\sqrt{A}}\,e^{-\hub\, \tau_0}\rt}
\lt \sqrt{A} + \frac{3}{4}\, i\, \hub\,
e^{\hub\, \tau_0}\rt
\ ,
\nonumber
\eeq
and we can set
\beq
c_1 = c_2 = 0
\ .
\eeq
Finally, the solution to the Pinney equation~\eqref{Pinney}
fulfilling these initial conditions is
\beq 
\sigma
\es 
\sqrt{\frac{\pi }{2\,\hub}} \,
\sqrt{ J^2_{\frac{\sqrt{B}}{\hub}}
\lt\frac{\sqrt{A}}{\hub} e^{-\hub\, \tau}\rt+
N^2_{\frac{\sqrt{B}}{\hub}}\left(\frac{\sqrt{A}}{\hub} 
e^{-\hub\,\tau}\right)}
\ .
\nonumber
\\
\eeq
\par
For a conformally coupled ($\xi=1/6$) massless ($\mu=0$)
scalar field in spatially flat ($K=0$) de~Sitter, one has $W_{1/6}=v$
[so that Eq.~\eqref{W/W} hold] and
\beq
\omega^2_{1/6}=
\omega^2 \lqu 1- \lt \frac{\hub\,a}{k} \rt^2 \rqu
\ ,
\label{ome16}
\eeq
which is real by virtue of the condition~\eqref{dSh}.
The solution to the Pinney
equation~\eqref{Pinney} then turns out to be very simply
\beq
\sigma(\tau) 
=\frac{1}{\sqrt{\omega}}
= \sqrt{\frac{a}{k}}
\ .
\label{sig_sol}
\eeq
The invariant annihilation operator~\eqref{bgen} becomes
\beq
\hat{b}
=
\frac{1}{\sqrt{2\, \hbar}} \lt \sqrt{k}\, a\, \pop
+ i\,\frac{\ppop}{\sqrt{k}\, a} \rt
\ ,
\eeq
from which
\beq
\begin{array}{l}
\pop
=\strut\displaystyle
\sqrt{\frac{\hbar}{2\,k\,a^2}} 
\lt \hat{b}^\dagger + \hat{b} \rt
\\
\\
\ppop
=
\strut\displaystyle i \, \sqrt{\frac{\hbar\,k\,a^2}{2}} 
\lt \hat{b}^\dagger - \hat{b} \rt
\ .
\end{array}
\eeq
\par
We also obtain the following useful relations
\beq
\lt \pop\,\ppop - \expec{ \pop\, \ppop} \rt \ket{n}=
\frac{i}{2}\, \hbar \lqu (\hat{b}^\dagger)^2 - \hat{b}^2 \rqu
\ket{n}
\label{refstart}
\eeq
\beq
&&
\lt \expec{ \pop^2\, \ppop^2} -\pop^2\, \ppop^2\rt
\ket{n}
\nonumber
\\
&&
\quad
=
\frac{\hbar^2}{4}
\lqu \hat{b}^4 + (\hat{b}^\dagger)^4  -4\,\bi^2 +4 \lt \bid \rt^2 \rqu 
\ket{n}
\eeq
\beq
\lt \hc - \expec{ \hc} \rt \ket{n}
=\frac{i}{2}\, \hbar\, \hub \lt \bi^2 - (\bid)^2 \rt
\ket{n}
\eeq
\beq
&&
\!\!\!\!\!\!\!\!\!
\lt \pop\, \ppop \,\ptau - \expec{ \pop\, \ppop\, \ptau} \rt
\!\ket{n}
=
\frac{i}{4}\,\hbar\,\hub
\left\{
\bi^2 - (\bid)^2
\phantom{\frac{A}{B}}
\right.
\nonumber
\\
&&
\!\!\!\!\!\!\!\!\!
\left.
- (\bid)^4- \bi^4 
-\frac{i\,k}{a\,\hub}
\lt 2\,n+1\rt \lqu (\bid)^2 - \bi^2 \rqu
\right\}
\ket{n}
\eeq
\beq
&&
\lt\ptau^2 -\expec{\ptau^2}\rt \ket{n} 
=
-i\, \frac{k\, \hub}{2\,a} 
\lqu
\lt 2\,n+1 \rt \lt \bi^2 - (\bid)^2 \rt
\right.
\nonumber
\\
&&
\quad
\left.
+i\,\frac{\hub\,a}{2\,k} \lt\bi^4+(\bid)^4 \rt \rqu
\ket{n}
\ .
\label{refend}
\eeq
Using Eq.~\eqref{sig_sol}, one can also write the coefficients
of the Bogoliubov transformation,
\beq
\begin{array}{l}
\hat{b} 
=
\zeta^*\, \hat{a}  + \eta\, \hat{a}^\dagger
\\
\\
\hat{b}^\dagger 
=
\zeta\, \hat{a}^\dagger  + \eta^*\, \hat{a}
\ ,
\end{array}
\eeq
as
\beq
\begin{array}{l}
\eta 
=
\strut\displaystyle\frac{1+i\,x -\sqrt{1-x^2}}{2 \lt 1-x^2\rt^{1/4}} 
\\
\\
\zeta
=
\strut\displaystyle\frac{1-i\,x +\sqrt{1-x^2}}{2 \lt 1-x^2\rt^{1/4}} 
\ ,
\end{array}
\eeq
where $x=\hub\,a/k$.
\null
\par
\null
\par
\null
\par
\null
\par
\null
\par
\null
\par
\null
\par
\null
\par
\null
\par
\null
\par
\null
\par
\null
\par
\null
\par
\null
\par
\null
\par
\null
\par
\null
\par
\null
\par
\end{document}